\documentstyle[12pt,epsf]{article}
\newcommand{\be}{\begin{equation}}
\newcommand{\ee}{\end{equation}}
\newcommand{\bc}{\begin{center}}
\newcommand{\ec}{\end{center}}

\begin{document}
\begin{flushright}
ITEP--PH--1--2000
\end{flushright}
\vspace{0.5cm}

\bc
{\large \bf The $\Lambda_b$ LIFETIME IN THE LIGHT FRONT 
QUARK MODEL.}\\ 
\vspace{1cm} 

{\bf P.Yu.Kulikov,
I.M.Narodetskii, A.I.Onischenko}
\ec
\bc
{\it Institute for Theoretical and Experimental Physics,\\ Moscow,
117218 Russia}.
\ec
\vspace{2.5cm}

\bc
{\Large\bf Abstract}
\ec
\vspace{0.5cm}

\noindent The enhancement of the $\Lambda_b$ decay width relative 
to $B$ decay one  
due to the difference of Fermi motion effects in $\Lambda_b$ and $B$ 
is calculated in 
%of the $b$and the lifetime of $\Lambda_b$ 
%baryon are analyzed within 
the 
light--front quark model with the simplifying assumption that $\Lambda_b$ 
consists of the heavy quark and light scalar diquark. 
%In addition we have 
%included the relatively small 
%corrections due to the nonspectator effects on $\tau(\Lambda_b)$. 
In order to explain the large 
deviation from unity in the experimental result for $\tau(\Lambda_b)/\tau(B)$, 
it is necessary that diquark be light and the ratio of the squares of the
$\Lambda_b$ and $B$ wave functions at the origin be $\le 1$.\\

\vspace*{2cm}

\noindent The lifetimes of the $b$ flavoured hadrons $H_b$ are related both to the CKM matrix
elements $|V_{cb}|$ and $|V_{ub}|$ and to dynamics of $H_b$ decays.
In the limit $m_b\to \infty$ the light quarks do not affect the decay of the
heavy quark, and thus the lifetimes of all $b$ hadrons must be equal.
The account of the soft degrees of freedom generates the preasymptotic
corrections which, however, have non-significant impact on the lifetimes and
various branching fractions of $B$ and $B_s$ mesons. Inclusive $H_b$ decays 
can be treated with the help of an operator product expansion (OPE) combined with 
the heavy quark expansion \cite{BSU97}. The OPE approach predicts that all corrections to 
the leading QCD improved parton terms appear at the order $1/m_b^2$ and beyond. 
Thus mesons and baryons containing b quark are expected to have lifetimes differing by no 
more than a few percent.  
The result of this approach for the $\Lambda_b$ lifetime is puzzling because it 
predicts that 
$\left(\tau(\Lambda_b)/\tau(B)\right)_{OPE}=0.98+{\cal O}(1/m^3_b)$ \cite{NS97}, 
whereas the 
experimental findings suggest a very much reduced fraction 
$\left(\tau(\Lambda_b)/\tau(B)\right)_{exp}
=0.78 \pm 0.04$ \cite{J97} or conversely 
a very much enhanced decay rate. The decay rates of B and 
$\Lambda_b$ are $\Gamma(B)=0.63\pm 0.02~{\rm ps}^{-1}$  and 
$\Gamma(\Lambda_b)=0.83\pm~0.05~{\rm ps}^{-1}$ differing by
$\Delta\Gamma(\Lambda_b)=
0.20\pm~0.05~{\rm ps}^{-1}$. The four--fermion processes of weak scattering and 
Pauli interference could  explain, under certain conditions, only $(13\pm 7)\%$ of this difference 
\cite{R96} (see, however, \cite {NS97}, \cite{DSM99}). 
In spite of great efforts of experimental activity the $\Lambda_b$ lifetime remains 
significantly low which continues to spur theoretical activity. 
In this respect, the use of
phenomenological models, like the constituent quark model, could be of
interest as a complementary approach to the OPE resummation method.

In this paper we shall compute the preasymptotic effects for the 
$\Lambda_b$ lifetime in the framework of the 
light--front (LF) quark model, which is a relativistic constituent quark 
model based on the LF formalism. In Ref. \cite{KNST99} this formalism has been 
used  to establish a simple quantum mechanical relation between the inclusive semileptonic decay 
rate of the B meson and that of a free ${\rm b}$ quark. The approach of \cite
{KNST99} relies on the idea of duality in summing over the final hadronic states. 
It has been assumed that the sum over all possible charm final states $X_c$ can be 
modelled by the decay width of an on--shell $b$ quark into on--shell $c$ quark 
folded with the $b$--quark distribution function 
$f_B^b(x,p^2_{\bot})=|\varphi_B^b(x,p^2_{\bot})|^2$
The latter represents the probability to find $b$ quark carrying 
a LF fraction $x$ of the hadron momentum and a transverse relative momentum 
squared $p^2_{\bot}$. For the semileptonic rates the abovementioned relation 
takes the form
\be
    \label{1}
    {d\Gamma_{SL}(B) \over dt} = {d\Gamma^b_{SL} \over dt}R_B(t),
\ee
where $d\Gamma^b_{SL}/dt$ is the free quark differential decay rate,
$t=q^2/m_b^2$, $q$ being the 4--momentum of the $W$ boson, and 
$R_B(t)$ incorporates the nonperturbative
effects related to the Fermi motion of the heavy quark inside the hadron. 
The expression for $d\Gamma^b_{SL}/dt$ for the case of 
non--vanishing lepton masses is given {\it e.g.} in \cite{KNST99}. 
$R(t)$ in (\ref{1}) is obtained by integrating
the bound--state factor $\omega(t,s)$ over the allowed region of  
the invariant hadronic mass $M_{X_c}$: 
\be
\label{2}
R_B(t)=    \int\limits_{s_{min}}^{s_{max}}ds \omega(t,s),
\end{equation}
where $s=M_{X_c}^2/m_b^2$ and 
%$\omega(t,s)$ is expressed in terms of  $f_B^b(x,p^2_{\bot})$ 
\begin{equation}
    \label{3}
    \omega(t,s) = m_b^2 x_0 ~ {\pi m_b \over q^+} ~ {|\bf{q}|
    \over |\bf{\tilde{q}}|} ~ \int\limits_{x_1}^{min[1, x_2]} dx ~
    |\varphi_B^b(x, p_{\perp}^{*2})|^2.
\end{equation}
In Eq. (\ref{3}) $x_0 = m_b/M_B$,
$p_{\bot}^{*2}=m_b^2(\xi(1-\rho-t)-\xi^2t-1)$ with
$\xi=\frac{xM_B}{q^+}$, and $\rho = (m_c / m_b)^2$, and the limits of
integration $x_{1,2}$ are given by $x_{1, 2} = x_0 q^+ / \tilde{q}^{\pm}$.  
%x_0 (q_0 +|{\bf q}|) /(\tilde{q}_0 \pm  |\bf{\tilde{q}}|)$.  
The plus component $q^+ = q_0 + |\bf q|$ is defined in the $B$ meson rest
frame whereas $\tilde q^{\pm} = \tilde q_0 \pm |{\bf \tilde q}|$ are defined 
in the $b$ quark rest frame. In Eq. (\ref{2}) the region of
integration over $s$ is defined through the condition $x_1 \le min[1,
x_2]$, {\it i.e.}
$s_{max} = x_0^{-2} (1 - x_0 \sqrt{t})^2$. 
For other details see \cite{KNST99}.

In the quark model the Fermi motion effect is due to 
the interaction with valence quark. The LF wave function 
$\varphi_B^b(x,p^2_{\bot})$ is defined in terms of the equal 
time radial wave function $\psi_B(p^2)$ as  \cite{C92} 
$\varphi^b_B(x,p^2_{\bot})=\frac{\partial p_z}{\partial x}
\frac{\psi_B(p^2)}{\sqrt{4\pi}}$, where $p^2=p^2_{\bot}+p^2_z$, 
$p_z=\left(x-\frac{1}{2}\right)M_0+\frac{m_{sp}^2-m_b^2}{2M_0}$, 
$M_0=\sqrt{p^2+m_b^2}+\sqrt{p^2+m_{sp}^2}$, and $m_{sp}$ is the 
constituent mass of the spectator quark. Explicit expression for 
$\frac{\partial p_z}{\partial x}$ can be found in \cite{GNST97}.
%\be
%\label{5}
%\varphi^b_B(x,p^2_{\bot})=
%\sqrt {\frac{M_0}{4x(1-x)}\left[1-
%\left(\frac{m_b^2-m^2_{sp}}
%{M_0^2}\right)^2\right]}\
%frac{\psi_B(p^2)}{\sqrt{4\pi}},
%\ee
%\be
%\label{6}
%\ee
%\noindent 
In what follows the B meson orbital wave function 
is assumed to be the Gaussian function as
\be
\label{7}
\psi_B(p^2)=\left(\frac{1}{\beta_{b\bar d}\sqrt{\pi}}\right)^{\frac{3}{2}}
\exp\left(-\frac{p^2}{2\beta^2_{b\bar d}}\right),
\ee
where the parameter $1/\beta_{b\bar d}$ 
defines the confinement scale. We take  $\beta_{b\bar d}=0.45~{\rm GeV}$ 
that is very close to the variational parameter $0.43~{\rm GeV}$ found in the  
Isgur--Scora model \cite{SI95} and corresponds to value of the QCD parameter 
$-\lambda_1=<p^2_b>=0.2~{\rm GeV^2}$. For $|\Psi_B(0)|^2$, the square of the 
wave function at the origin, 
we have $|\Psi_B(0)|^2=1.64\cdot 10^{-2}~{\rm GeV^3}$. This value compares favourably 
with the estimation in the constituent quark ans\"atz \cite {S82} 
$|\Psi_B(0)|^2=M_Bf^2_B/12=(1.6\pm 0.7)\cdot 10^{-2}~{\rm GeV}^3$ for 
$f_B=190\pm 40~{\rm MeV}$. 

The same formulae can be also applied for nonleptonic 
$B$ decay widths (corresponding to the underlying quark decays 
$b\to cq_1q_2$) thus making it possible to calculate the $B$ 
lifetime \cite {KN00}. The lepton pair is substituted by a quark 
pair, so that  
$d\Gamma_{SL}^b/dq^2$ is replaced by
$d\Gamma_{NL}^b/dq^2=\eta |V_{q_1q_2}|^2
d\Gamma_{SL}^b/dq^2$, where (in the limit $N_c\to \infty$) 
$\eta=\frac{3}{2}(c_+^2+c_-^2)$, 
%$c_{1,2}=\frac{1}{2}(c_+\pm c_-)$ 
with $c_-$ and $c_+=c_-^{-1/2}$ being the standard short distance QCD 
enhancement and suppression factors in a color antitriplet and sextet, 
respectively. We take $c_+=0.84$, $c_-=1.42$. 

The constituent quark masses are the free parameters in our model. 
We have found that the $\tau_{\Lambda_b}/\tau_B$ ratio is  rather stable 
with respect to the precise values 
of the heavy quark masses $m_b$ and $m_c$ provided $m_b-m_c\ge 3.5~{\rm GeV}$. From now on we 
shall use the reference values $m_b=5.1~{\rm GeV}$ and $m_c=1.5~{\rm GeV}$. 
The value of the CKM parameter 
$|V_{cb}|$ 
cancels in the ratio $\tau_{\Lambda_b}/\tau_{B}$, but is important 
for the absolute rates. Details of our calculations of $\Gamma(B)$ 
are given in Table 1 for the three different values of the 
constituent mass $m_{sp}$. The values $m_{sp}\sim 300~(200)~{\rm MeV}$ are usually 
used in non-relativistic (relativized) quark models \cite{SI95},\cite{GI85}. 
We have also 
considered a very low constituent quark mass $m_{sp}=100~{\rm MeV}$ to see how 
much we can push up the theoretical prediction of the $\Gamma(\Lambda_b)
/\Gamma(B)$, see below. All the semileptonic widths include 
the pQCD correction as an overall reduction factor equal to 0.9.  
%$|V_{cb}|=0.040$ we obtain for the total rate of the B meson 
%$\Gamma_B=0.00~{\rm ps}^{-1}$ ($\tau_B=1.56~{\rm ps}$). 
Following Ref. \cite{GNST97} we have included the transitions to baryon-antibaryon 
$(\Lambda_c\bar N~{\rm and}~\Xi_{cs}\bar \Lambda)$ pairs. 
In addition we have added ${\rm BR}\approx 1.5\%$ for the 
Cabbibo--suppressed $b\to u$ decays with $|V_{ub}/V_{cb}|\sim 0.1$.  
The value of 
$|V_{cb}|$ is defined by the condition that the calculated B lifetime 
is $1.56~ps$.
 
Now we turn to the calculation of the $\Lambda_b$ decay rate. We shall analyze the inclusive semileptonic and 
non--leptonic  $\Lambda_b$ rates on the simplifying asumption that $\Lambda_b$ is composed 
of a heavy quark and a light scalar diquark with the effecive mass $m_{ud}$.
Then the treatment of the inclusive $\Lambda_b$ decays is simlified to a 
great extent and we can apply the model considered above with the minor modifications. 
For the heavy--light diquark wave function $\psi_{\Lambda_b}$ 
we again assume the Gaussian ans\"atz with the 
oscillator parameter $\beta_{bu}$. The width of $\Lambda_b$ can be obtained from that 
of $B$ by the replacements 
$M_B\to M_{\Lambda_b}$, $m_{sp}\to m_{ud}$ and $\beta_{b\bar d}\to \beta_{bu}$. 
Note that the latter two replacements change $f_{\Lambda_b}^b$, 
the $b$ quark distribution function inside the $\Lambda_b$, in comparison with $f_B^b$.

The inclusive nonleptonic channels for $\Lambda_b$ are the same as for B meson 
except for the decays into baryon--antibaryon pairs which are missing in case 
of $\Lambda_b$. The absence of this decay channel leads to the reduction of 
$\Gamma(\Lambda_b)$ by $\approx 7\%$. This reduction can not be compensated 
by the phase space enhancement in $\Lambda_b$. 
%enhancement due 
%to the four--quark operators \cite{R96}. The phase space enhancement in 
%$\Lambda_b$ decays is marginal and can not be responsible for a shorter lifetime 
%of $\Lambda_b$. 
The only way to get an enchancement 
of $\Gamma_{\Lambda_b}$ is to enhance its non--leptonic rates. Altarelly 
{\it et al.} \cite{AMPR96} have suggested the increase of the non--leptonic 
rates could 
be due to the phenomenological factor $(M_{H_b}/m_b)^5$, then the $6\%$ 
difference between $M_{\Lambda_b}$ and $M_B$ is enough to explain the 
experimentally observed enhancement. In our approach, the only distinction 
between the two lifetimes, $\tau_{\Lambda_b}$ and $\tau_B$, 
can occur  due to the difference of Fermi motion effects encoded in 
$f_{\Lambda_b}^b$ and $f_B^b$.

The $f_{\Lambda_b}^b$ is defined by the two parameters,  
$m_{ud}$ and $\beta_{ub}$.  The later quantity can be translated into the ratio 
of the squares of the wave functions determining the probability to find a light 
quark at the location of the $b$ quark inside the $\Lambda_b$ baryon and $B$ 
meson, {\it i.e.}
\be
\label{8}
r=\frac{|\Psi_{\Lambda_b}(0)|^2}{|\Psi_B(0)|^2}=
\left(\frac{\beta_{bu}}{\beta_{b\bar d}}\right)^3.
\ee
Estimates of the parameter $r$ using the non--relativistic quark model or 
the bag model  \cite{V85}, \cite{BGT84}, \cite{B94} or QCD sum rules 
\cite{CF96} are typically in the range $0.1-0.5$. 
On the other hand, 
Rosner has estimated the heavy--light diquark density at zero separation 
in $\Lambda_b$ from the ratio of hyperfine splittings between $\Sigma_b$ and
 $\Sigma_b^*$ baryons and $B$ and $B^*$ mesons and finds \cite{R96}
\be
\label{9} 
r=\frac{4}{3}\cdot \frac{
m_{\Sigma^*_b}^2-m_{\Sigma^*_b}^2}{
m_{B^*}^2-m_B^2}.
\ee
This lead to $r\sim 0.9\pm 0.1$, if the baryon splitting is taken to be 
$m_{\Sigma^*_b}^2-m_{\Sigma_b}^2
\sim m_{\Sigma^*_c}^2-m_{\Sigma_c}^2=
(0.384\pm 0.035)~{\rm GeV}^2$, or even to $r\sim 1.8\pm 0.5$, if the surprisingly 
small and not confirmed yet DELPHI result 
$m_{\Sigma^*_b}-m_{\Sigma_b}=(56\pm 16)~{\rm MeV}$ \cite{D95} is used.

On the other hand, the width $\Gamma(\Lambda_b)$ and hence the ratio 
$\tau(\Lambda_b)/\tau(B)$ is very sensitive to 
the choice of $m_{ud}$ and $r$. In order to study the dependence on 
$m_{ud}$ and $r$ we keep the values of the quark masses $m_b$ and $m_c$ 
fixed and vary the wave 
function ratio in the range $0.3\le r\le 2.3$ that corresponds to 
$0.3~{\rm GeV}\le \beta_{bu}\le 0.6~{\rm GeV}$. We take two representative 
values for the diquark mass: $m_{ud}=m_u+m_d$ 
corresponding to zero binding approximation and 
$m_{ud}=m_*\approx \frac{1}{2}(m_u+m_d-m_{\pi})$. In the latter relation 
inspired by the quark model, 
the factor $1/2$ arises from the diferent color factors for $u$ and $\bar d$ 
in the $\pi$--meson  ( a triplet and antitriplet making a singlet) and 
$u$ and $d$ in the the $\Lambda_b$ (two triplets making an antitriplet).

%The diference between $\Gamma(\Lambda_b)$ and $\Gamma(B)$  
%comes almost entirely from the difference between distribution functions 
%$f_{\Lambda_b}^b(x,p^2_{\bot})$. 
In Fig. 1 we compare one--dimensional distribution 
functions 
\be
\label{10}
F_{\Lambda_b}^b(x)=\pi\int\limits_0^{\infty} dp^2_{\bot}f_{\Lambda_b}^
b(x,p^2_{\bot})
\ee 
with that of the B meson. 
These functions exhibit a pronounced maximum at 
$m_b/(m_b+m_{sp})$ (in case of the $B$ meson) and $m_b/(m_b+m_{ud})$ 
(in case of $\Lambda_b$). The width of $F_{\Lambda_b}$ depends on $\beta_{bu}$ 
and goes to zero when $\beta_{bu}\to 0$. 
Note that the calculated branching fractions of $\Lambda_b$  show marginal 
dependence on the choice of the model parameters; they are $\sim 11.5~\%$ 
for the semileptonic $b\to ce\nu_e$ transitions, 
$\sim 2.8\%$ for  $b\to c\tau\nu_{\tau}$, $\sim 50\%$ for the nonleptonic 
$b\to cd\bar u$ transitions, and $\sim 16\%$ for $b\to c\bar cs$ 
transitions.

Our results for 
$\tau_{\Lambda_b}/\tau_{B}$ 
are shown in Table 2 and 
Fig. 2. We have also included the contribution of four-quark 
operators calculated using the factorization approach and the description 
of the baryon relying on quantum mechanics of only the constituent quarks. 
This contribution leads to a small enhancement 
of the $\Lambda_b$ decay rate by an amount
\be
\label{r}
\Delta\Gamma^{4q}(\Lambda_b)=\frac{G_F^2}{2\pi}|\Psi_{bu}(0)|^2|V_{ud}|^2
|V_{cb}|^2m_b^2(1-\rho)^2[c^2_--(1+\rho)c_+(c_--c_+/2)].
\ee
This contribution scales like $\beta_{bu}^3$ and 
varies between $0.01
~{\rm ps}^{-1}$ and $0.03~{\rm ps}^{-1}$ when $\beta_{bu}$ varies 
between 0.35 and 0.55 GeV.

The quantity $\tau_{\Lambda_b}/\tau_{B}$ is particular sensitive to the light 
quark mass $m_{sp}$. We observe that to decrease the theoretical prediction for 
$\tau_{\Lambda_b}$ requires to decrease the value of the hadronic parameter 
$r$ in (\ref{8}) to 0.3-0.5 and the value of $m_{sp}$ to $\sim 100~{\rm MeV}$. 
For example, assuming that $r\sim 0.3$ we find 
that  the lifetime ratio is decreased from $0.88$ to $0.81$ if 
$m_{sp}$ is reduced from $300~{\rm MeV}$ to $100~{\rm MeV}$ and the diquark 
mass is chosen as $m_{ud}=m_u+m_d$. For the diquark mass 
$m_{ud}\sim m_*$
%\approx\frac{1}{2}(m_u+m_d-m_{\pi} $ 
the ratio is almost stable ($\sim  
0.8$), so that reducing of the diquark mass produces a decrease 
of the lifetime ratio by $1\%$, $5\%$, and $8\%$ for $m_{sp}=100,~200,~
{\rm and}~300~{\rm MeV}$, respectively. 
Varying the spectator quark mass in a similar way we find that for 
the "central value" $r\sim 1$ the lifetime ratios 
are reduced from $0.93$ to 0.88 for $m_{ud}=m_u+m_d$ and remain almost 
stable ($\sim 0.86$) for $m_{ud}\sim m_*$.  
For the largest possible value of $r$ suggested in \cite{R96}, 
$r\sim 2.3$, the lifetime ratios 
are reduced from 0.97 to 0.94 in the 
former case and remain almost stable $\sim 0.91-0.93$ in the latter 
case. 

If the current value of 
$\left(\tau(\Lambda_b)/\tau(B)\right)_{exp}$ persists, the most likely its 
explanation is that some hadronic matrix elements of four--quark operators 
are larger than the naive expectation (\ref{r}) \cite{NS97}. 
A recent lattice study of Ref. \cite{DSM99} suggests that the effects of weak 
scattering and interference can be pushed at the $\approx 8\%$ level for 
$r=1.2\pm 0.2$ {\it i.e.}  for the value of $r$ that is significantly larger 
that most quark model predictions but smaller than the upper Rosner estimation.  
If a significant fraction $\sim50\%$ of the discrepancy between the theoretical 
prediction for $\tau_{\Lambda_b}/\tau_{B}$ and the experimental 
result can be accounted for the spectator effects then 
the reminder of the discrepancy can be explained by the preasymptotic 
effect due to Fermi motion of the $b$ quark inside $\Lambda_b$. Indeed, choosing the quite reasonable values of the spectator 
quark mass $m_{sp}=200~{\rm MeV}$ and the diquark mass 
$m_{ud}=250~{\rm MeV}$ and subtracting the small contribution (\ref{r}) we 
find that the Fermi motion effect produces for $r= 1.2\pm 0.2$ an additional 
reduction of $\tau_{\Lambda_b}$ by $12\pm 2\%$.

\vspace{0.2cm}

\noindent This work was supported in part by INTAS grant Ref. 
no 96--155 and RFBR grants Refs. no 00-02-16363 and 00-15-96786.

\vspace{2cm}

\newpage
\noindent {\bf Table 1.} The branching fractions (in per cent) for the
inclusive semileptonic 
and nonleptonic B decays calculated within the LF quark model for the 
several values of the constituent quark mass $m_{sp}$. The heavy quark masses 
are $m_b=5.1~{\rm GeV}$, $m_c=1.5~{\rm GeV}$. The oscillator parameter in Eq. 
(\ref{7}) is $\beta_{b\bar d}=0.45~{\rm GeV}$. The values of $|V_{cb}|$ in units of 
$10^{-3}\sqrt{1.56~{\rm ps}/\tau^{(exp)}(B)}$ are also reported.
\vspace{0.2cm}

\begin{table}[th]
\begin{center}
\begin{tabular}{|c|c|c|c|}
\hline\hline & $m_{sp} = 100~{\rm MeV}$ & $m_{sp} = 200~
{\rm MeV}$ & $m_{sp} = 300~{\rm MeV}$ \\ \hline\hline $b\to ce\nu_e $
& 10.65 & 10.98 & 11.46 \\ $b\to c\mu\nu_{\mu} $ & 10.59 & 10.93 &
11.40 \\ $b\to c\tau\nu_{\tau} $ & ~2.47 & ~2.51 & ~2.57 \\ \hline
$b\to cd\bar u $ & 47.88 & 47.88 & 47.52 \\ $b\to c{\bar c}s $ &
14.07 & 14.31 & 14.63 \\  
$b\to cs\bar u$ & ~2.94 & ~3.09 & ~3.32 \\ \hline
$B\to \Xi_{cs}\bar \Lambda_c $ & ~2.22 & ~1.83
& ~1.43 \\ $B\to \Lambda_c \bar N $ & ~7.70 & ~6.91 & ~6.02 \\ \hline 
$b\to u $ & ~1.47 & ~1.54 &
~1.66 \\ \hline\hline $|V_{bc}|$ & 38.3 & 39.3 & 40.7 \\ \hline\hline 
%$\Gamma_{tot}/\Gamma_0$
%(\%) & 101.95 & 97.02 & 90.42 \\ \hline
\end{tabular}
\end{center}
%\caption{ } 
%\label{Branchings}
\end{table}
\vspace{0.3cm}

\noindent {\bf Table 2}. The LF quark model results for the ratio 
$\tau_{\Lambda_b}/\tau_{B}$ calculated for different sets of paremeters. 
The diquark masses are given in units of MeV, whereas $\beta_{bu}$ are
in units of GeV.

\vspace{0.2cm}

\begin{table}[th]
\begin{center}
\begin{tabular}{|c|c|c|c|}
\hline\hline   & $m_{sp} = 100~\mbox{MeV}$ & $m_{sp} = 200~\mbox{MeV}$ &
$m_{sp} = 300~\mbox{MeV}$
\\ \hline
 $\beta_{bu}
\setminus m_{ud}$ & \begin{tabular}{c|c} ~~200 & ~~150 
\end{tabular} &
 \begin{tabular}{c|c} ~~400 & ~~250 \end{tabular} & 
\begin{tabular}{c|c} ~~600 & ~~400 \end{tabular}
 \\ \hline\hline
0.3 & \begin{tabular}{c|c} 0.807 & 0.795
\end{tabular} &
 \begin{tabular}{c|c} 0.843 & 0.792 \end{tabular} & \begin{tabular}{c|c} 0.885 & 0.803 \end{tabular}
 \\ \hline
0.45 & \begin{tabular}{c|c} 0.877 & 0.867
\end{tabular} &
 \begin{tabular}{c|c} 0.901 & 0.857 \end{tabular} & \begin{tabular}{c|c} 0.932 & 0.859 \end{tabular}
 \\ \hline
0.6 & \begin{tabular}{c|c} 0.937 & 0.929
\end{tabular} &
 \begin{tabular}{c|c} 0.951 & 0.913 \end{tabular} & \begin{tabular}{c|c} 0.970 & 0.906 \end{tabular}
 \\ \hline\hline
\end{tabular}
\end{center}
%\caption{} \label{Ratio}
\end{table}
%\vspace{1cm}

\section*{Figure captions}
%\vspace{0.5cm}
\noindent {\bf Figure 1}. The distribution functions $F_B^b(x)$ (solid line) 
and $F_{\Lambda_b}^b(x)$ (thin lines) defined by Eq. (\ref{10}) versus 
the LF momentum fraction $x$. Labels 1 to 4 
on the curves refer to the cases 
$\beta_{bu}=0.3~{\rm GeV},~m_{ud}=600~{\rm MeV}$;
$\beta_{bu}=0.3~{\rm GeV},~m_{ud}=300~{\rm MeV}$;
$\beta_{bu}=0.6~{\rm GeV},~m_{ud}=600~{\rm MeV}$;
$\beta_{bu}=0.6~{\rm GeV},~m_{ud}=300~{\rm MeV}$, respectively.\\[1cm]
\noindent {\bf Figure 2}. The lifetime ratios 
$\tau_{\Lambda_b}/\tau_B$ for 
$\beta=0.3~{\rm GeV}~(r=0.3))$ (solid line), $\beta=0.45~{\rm GeV}~
(r=1)$ (long--dashed line), and 
$\beta=0.6~{\rm GeV}~(r=2.37)$ (short--dashed line). (a) $m_{ud}=m_u+m_d$, and (b) 
$m_{ud}\sim\frac{1}{2}(m_u+m_d-m_{\pi})$.

\newpage 
\begin{minipage}[t]{10.cm}
\begin{center}
\includegraphics{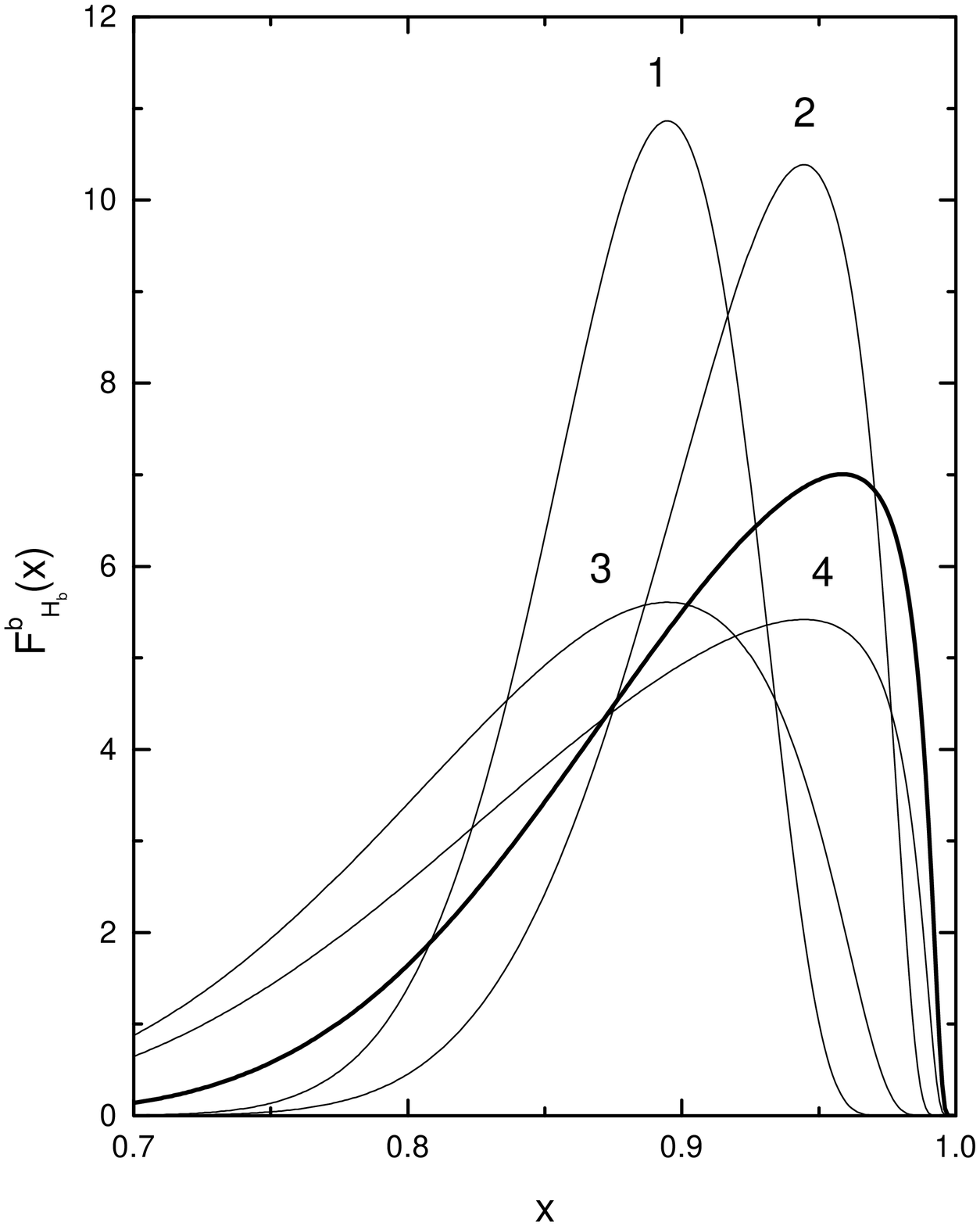}
\includegraphics{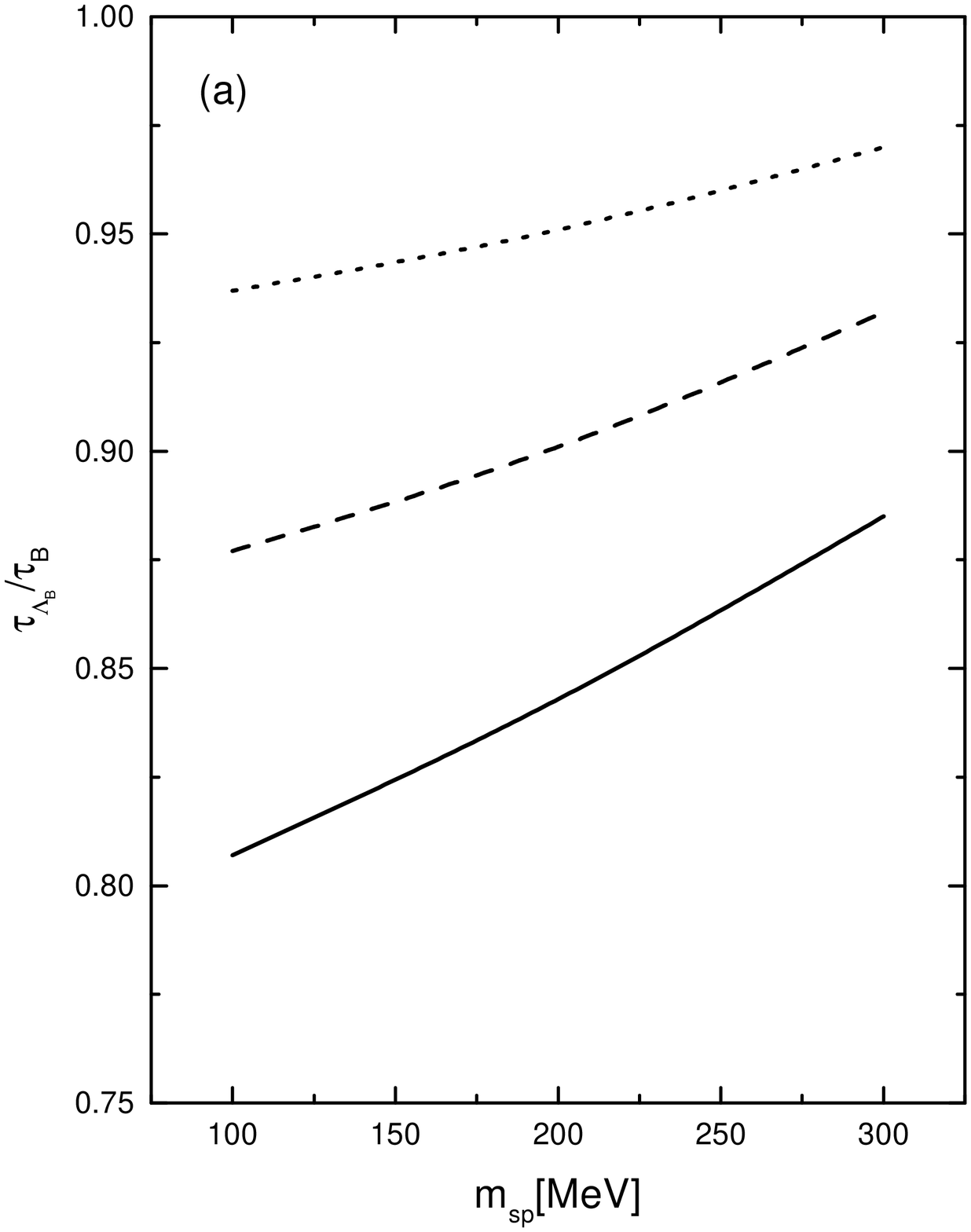}
\includegraphics{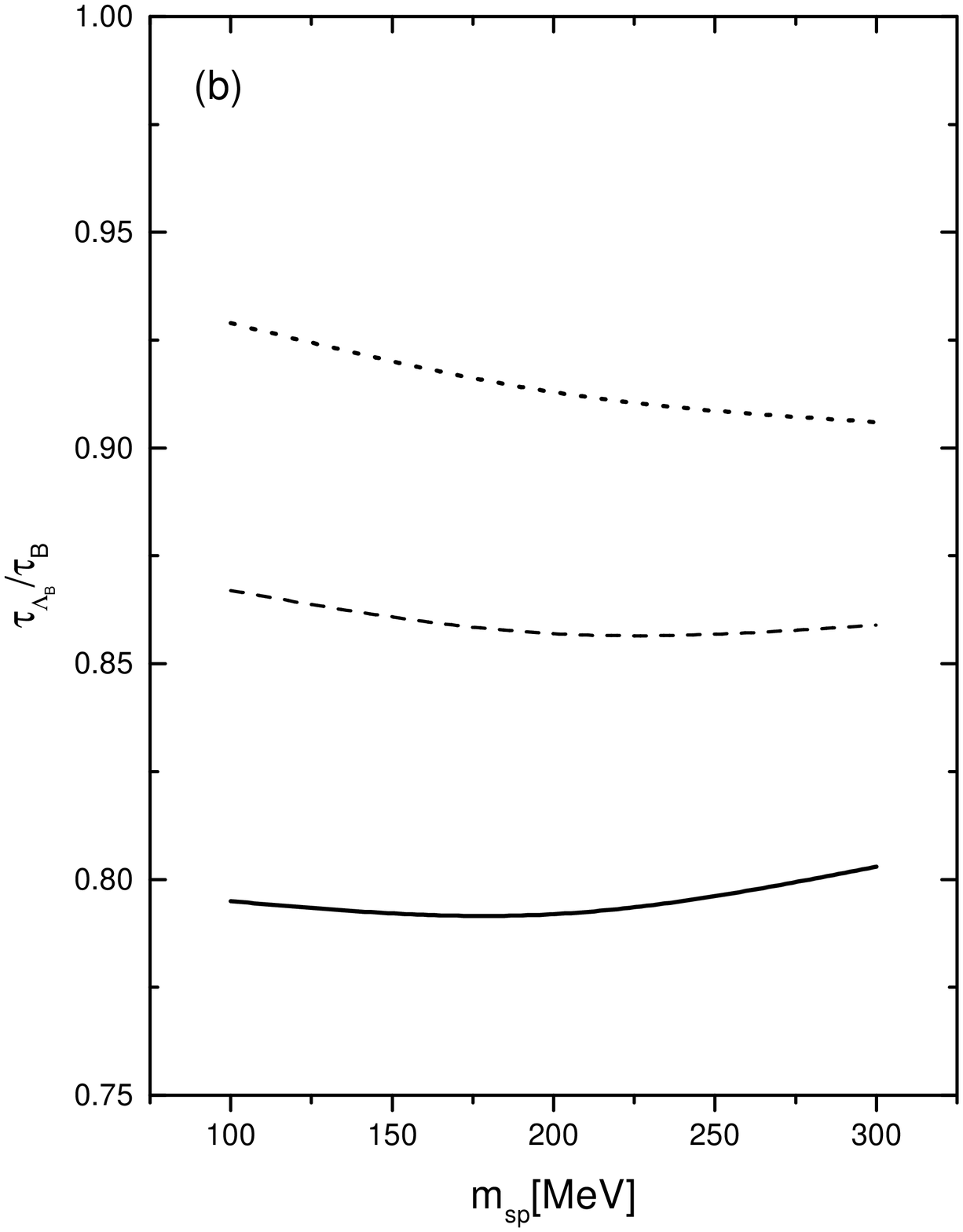}
\end{center}
\vspace*{0.8cm}
\end{minipage}
\vskip 8cm
\bc
{\bf Figure 1} 
\vskip 11cm
{\bf Figure 2} 
\ec
\end{document}